\documentclass[12pt]{article}
\usepackage{graphicx}
\usepackage{dcolumn}
\usepackage{bm}
\usepackage{graphics}
\usepackage{subfigure}
\begin{document}

\begin{center}

{\bf{\large Evolution Equations for Connected and Disconnected Sea Parton Distributions}}

\vspace{0.6cm}


{\bf  Keh-Fei Liu}

\end{center}

\begin{center}
\begin{flushleft}
{\it
Department of Physics and Astronomy, University of Kentucky, Lexington, KY 40506
}
\end{flushleft}
\end{center}

\begin{abstract}
It has been revealed from the path-integral formulation of the hadronic tensor that there are connected sea and disconnected sea partons. The former is responsible for the Gottfried sum rule violation primarily and evolves the same way as the valence. Therefore, the DGLAP evolution equations can be extended to accommodate them separately. We discuss its consequences and implications vis-a-vis
lattice calculations. 
\end{abstract}


\section{Introduction}

Partonic structure of the nucleon has been discovered and extensively studied in deep inelastic 
scattering (DIS) of leptons. Further experiments in Drell-Yan process, semi-inclusive DIS (SIDIS) help to identify and clarify the flavor dependence, particularly the sea partons~\cite{Peng:2014hta}.  The first experimental evidence that the sea patrons have non-trivial flavor dependence is revealed in the experimental demonstration of the violation of Gottfried sum rule. The original 
Gottfried sum rule, $I_G \equiv \int^1_0 dx [F^p_2(x)-F^n_2(x)]/x  =1/3$, was obtained
under the assumption that $\bar u$ and $\bar d$ sea partons are the same~\cite{gottfried}. However, the NMC measurement~\cite{nmc} 
of $\int^1_0 dx [F^p_2(x)-F^n_2(x)]/x$ turns out to be $0.235 \pm 0.026$, a 4 $\sigma$ difference from the Gottfried sum rule, which implies that the $\bar u = \bar d$ assumption was invalid. The correct expression for the Gottfried sum in the quark parton model should be~\cite{nmc}
\begin{equation}   \label{gsm}
I_{p-n} \equiv \int^1_0 \frac{[F^p_2(x)-F^n_2(x)]}{x}dx = \frac{1}{3} + \frac{2}{3} \int dx (\bar{u}(x) - \bar{d}(x)) + 
\mathcal{O}(\alpha_s^2) 
\end{equation}
so that the $x$-integrated difference of the $\bar u$ 
and $\bar d$ sea is \mbox{$\int^1_0 [\bar d(x) - \bar u(x)] dx =0.148 \pm 0.039$.} This striking result from the NMC was subsequently checked using an independent experimental technique. From measurements
of the Drell-Yan (DY) cross section ratios of $(p+d)/(p+p)$, the NA51~\cite{na51} and the Fermilab E866~\cite{e866} experiments clearly observed the $\bar u$ and $\bar d$ difference in the
proton sea over the kinematic range of $0.015 < x < 0.35$.

This came as a surprise at the time, because it was previously assumed that the sea partons originate from the gluon splitting
(i.e. $g \rightarrow u\bar{u}, d\bar{d},  s\bar{s}$) in a flavor-blind manner. Since the perturbative calculation leading to $\bar{u} - \bar{d}$ difference is at the two loop level which is too small to explain the size of the difference~\cite{Ross:1978xk,Broadhurst:2004jx}, it must have come from the intrinsic higher Fock-space wavefunction of the nucleon, e.g. $q^4\bar{q}$ component. Several meson cloud models~\cite{tony,kumano,garvey} have been used to 
explain this difference via the Sullivan process~\cite{Sullivan:1971kd}. The non-perturbative origin for such a difference is explained in QCD itself via the Euclidean path-integral formulation of the hadronic tensor~\cite{Liu:1993cv,Liu:1998um,Liu:1999ak}. It is shown that there are two kinds of sea partons -- the connected sea (CS) and disconnected sea (DS) partons and the Gottfried sum rule violation comes exclusively from the CS at the isospin symmetry limit~\cite{Liu:1993cv}. In view of this $\bar{u} - \bar{d}$ difference discovered in DIS and the similar finding in DY process of non-unity ratio of $\bar{u}(x)/\bar{d}(x)$, the global fitting have since accommodated this. However, the CS origin of $\bar{u} - \bar{d}$ has not be incorporated in the fitting of $\bar{u} + \bar{d}$ and it is not recognized that $\bar{u}$ and 
$\bar{d}$ have two origins, i.e. the CS and DS and only the DS part has the same small $x$ behavior as that 
of $\bar{s}$.  An attempt to separate the CS and DS parts of  $\bar{u} + \bar{d}$ has been carried out~\cite{Liu:2012ch} by combining the the CT10 global fitting with the HERMES data
on $s+\bar{s}$ and a lattice calculation of $\langle x\rangle_s / \langle x\rangle_{u/d}$ (DI) where 
$\langle x\rangle_{u/d}$ (DI)  is the momentum fraction of $u/d$ in the disconnected insertion (DI) calculation on the 
lattice~\cite{doi08}.

    To separate CS and DS $u$ and $d$ parton distributions and fitted to different experiments at different kinematics, they need to be evolved from one $Q^2$ to another. In this manuscript, we shall present the extended evolution equations which accommodate 
 differently evolved CS and DS. We shall start by reviewing the status-quo DGLAP evolution equations in Sec.~\ref{EE}. The 
formulation of the hadronic tensor in the path-integral formalism is given in Sec.~\ref{HT}. The classification of the parton degrees of freedom is given in Sec.~\ref{Pdegree} with an example of separating the CS from the DS
by combining results from SIDIS results of the strange parton distribution, the global fitting of the parton distribution function
(PDF) and the lattice calculation. The extended NNLO evolution equations accommodating CS and DS separately are
given in Sec.~\ref{CSDSEE}. Also included in Sec.~\ref{CSDSEE} are comments of their implications and their relation
to lattice calculations. Finally,  a summary is given in Sec.~\ref{sum}.

\section{NNLO Evolution Equations}  \label{EE}

To begin with, we shall review the present implementation of the NNLO evolution equations which starts with the following DGLAP equations~\cite{mvv04,ccg08,nad} with $t \equiv \ln \mu$

\begin{eqnarray}  \label{evo_eq_q}
\frac{dq_i}{dt}\!\! &=& \!\! \sum_k (P_{ik} \otimes q_k + P_{i\bar{k}}\otimes \bar{q}_k) + P_{ig} \otimes g; \\
\frac{d\bar{q}_i}{dt} \!\!&=& \!\! \sum_k (P_{\bar{i}k} \otimes q_k + P_{\bar{i}\bar{k}}\otimes \bar{q}_k) 
+ P_{\bar{i}g} \otimes g; \label{evo_eq_aq}\\
\frac{dg}{dt}\!\! &=& \!\! \sum_k (P_{gk} \otimes q_k + P_{g\bar{k}}\otimes \bar{q}_k) + P_{gg} \otimes g.
\end{eqnarray}
where the splitting function (kernel in the integral) P are~\cite{mvv04,ccg08}

\begin{eqnarray}
P_{ik}\!\!\! &= &\!\!\!P_{\bar{i}\bar{k}} = \delta_{ik} P_{qq}^v + P_{qq}^s, \label{Pik}, \hspace{0.5cm}
P_{\bar{i}k} = P_{i\bar{k}} = \delta_{ik} P_{q\bar{q}}^v + P_{q\bar{q}}^s, \label{Pibark} \\
P_{ig} \!\!\!& = & \!\!\!P_{\bar{i}g} \equiv P_{qg};  \hspace{2.2cm}
P_{gi} =  P_{g\bar{i}} \equiv P_{gq}.
\end{eqnarray}


The practical approach takes the following combinations of quark PDF's so that some of the combined PDF's
evolve independently.

\begin{eqnarray}
\Sigma\!\!\! & \equiv& \!\!\! \sum_{i} (q_i + \bar{q}_i); \hspace{1.1cm}
\Sigma_v \equiv \sum_{i} (q_i - \bar{q}_i);  \label{def_sigma}\\
q_i^+ \!\!\!\!&\equiv & \!\!\!  q_i + \bar{q}_i - \frac{1}{N_f} \Sigma; \hspace{0.5cm}
q_i^- \equiv q_i - \bar{q}_i.  \label{def_qpm}
\end{eqnarray}


The evolution equations are written in terms of these combined distributions

\begin{eqnarray}   
&&\frac{d\Sigma_v}{dt} =P_{vv} \otimes \Sigma_v;  \label{sigma_v}\\
&&\frac{dq_i^+}{dt} =  P_{qq}^+ \otimes q_i^+;  \label{q+}\\
&&\frac{dq_i^-}{dt} =  P_{qq}^- \otimes q_i^- + (P_{qq}^s - P_{q\bar{q}}^s) \otimes \Sigma_v,  \label{q-} \\
&&\frac{d\Sigma}{dt} =  P_{\Sigma\Sigma} \otimes \Sigma + P_{\Sigma g} \otimes g,  \label{sigma_sum}\\
&&\frac{dg}{dt}=  P_{g\Sigma} \otimes \Sigma + P_{gg} \otimes g, \label{glue}
\end{eqnarray}
with 
\begin{eqnarray}
P_{vv} \!\!\! &=& \!\!\! P_{qq}^v - P_{q\bar{q}}^v + N_f (P_{qq}^s - P_{q\bar{q}}^s); \\
P_{qq}^+ \!\!\! &=& \!\!\!  P_{qq}^v + P_{q\bar{q}}^v; \hspace{0.5cm}
P_{qq}^- = P_{qq}^v - P_{q\bar{q}}^v; \\
P_{\Sigma\Sigma} \!\!\! &=& \!\!\!  P_{qq}^v + P_{q\bar{q}}^v + N_f ( P_{qq}^s + P_{q\bar{q}}^s); \\
P_{\Sigma g} \!\!\! &=& \!\!\! 2 N_f P_{qg}; \hspace{0.5cm}
P_{g\Sigma} = P_{gq}.
\end{eqnarray}

    Notice that there is an inhomogeneous term $\Sigma_v$ in Eq.~(\ref{q-}) which is the sum of all flavors. Since $q_i^-$ has usually been defined as the valence quark by conventional wisdom, it seems to imply, on the surface, that a valence $u$ quark can evolve into 
a valence $d$ quark and vice versa. This is not possible in QCD, of course, since it does not have flavor-changing couplings.
To trace its origin, one can see that it comes from the $P_{qq}^s$ and $P_{q\bar{q}}^s$
terms in Eqs.~(\ref{Pik})  which are different. This is due to the exchange of three gluons between
the quark loop with current insertions and the quark line from the nucleon (valence or sea) as shown in
Fig. (1b) in Ref.~\cite{lv91}. This gives rise to a difference in the parton and anti-parton distributions in
the disconnected sea (i.e. in the quark loop) which is not valence. For example, $s^-(x) = s(x) -\bar{s}(x)$ is not
valence in the nucleon even though the net valence strangeness is zero, i.e. $\int dx\, s^-(x) = 0$. Therefore,
the definition $q_i^-$ is not valence in NNLO, since the parton-antiparton difference can be generated in the disconnected sea.
As we shall see later in Sec.~\ref{CSDSEE}, when we expand the evolution equations to separate out the valence, the connected sea and the disconnected sea, Eq.~(\ref{q-}) is actually a linear combination of two equations, one involves the valence and
and the connected sea and other the disconnected sea. This will help clarify the meaning and definition of $q^-(x)$.

\section{Hadronic tensor in path-integral formalism}  \label{HT}

The deep inelastic scattering of a muon on a nucleon involves the hadronic
tensor which, being an inclusive reaction, includes all intermediate states
\begin{equation}   \label{w}
W_{\mu\nu}(q^2, \nu) = \frac{1}{2} \sum_n \int \prod_{i =1}^n \left[\frac{d^3 p_i}{(2\pi)^3 2E_{pi}}\right]  \langle N|J_{\mu}(0)|n\rangle
\langle n|J_{\nu}(0) | N\rangle_{spin\,\, ave.}(2\pi)^3 \delta^4 (p_n - p - q) . 
\end{equation}

Since deep inelastic scattering measures the absorptive part of the  
Compton scattering, it is the imaginary part of the forward amplitude and
can be expressed as the current-current correlation function in the nucleon, 
i.e.
\begin{equation}  \label{wcc}
W_{\mu\nu}(q^2, \nu) = \frac{1}{\pi} {\rm Im} T_{\mu\nu}(q^2, \nu)
= \langle N| \int \frac{d^4x}{4\pi}  e^{ i q \cdot x} J_{\mu}(x)
J_{\nu}(0) | N\rangle_{spin\,\, ave.}.
\end{equation} 

It has been shown~\cite{Liu:1993cv,Liu:1998um,Liu:1999ak,Aglietti:1998mz,Detmold:2005gg,Liu:2016djw} that the 
hadronic tensor $W_{\mu\nu}(q^2, \nu)$ can be obtained from the Euclidean path-integral formalism. 
In this case, one considers the ratio of the four-point function  
\mbox{$\langle \chi_N(\vec{p},t_f) \int \frac{d^3x}{4\pi} 
e^{- i \vec{q}\cdot  \vec{x}} J_{\nu}(\vec{x},t_2) J_{\mu}(0,t_1)
\chi_N(\vec{p}, t_0)\rangle$} and the two-point function
\mbox{$\langle \chi_N(\vec{p},t_f) \chi_N(\vec{p},t_0)\rangle$},
where $\chi_N(\vec{p},t)$ is an interpolation
field for the nucleon with momentum $p$ at Euclidean time $t$.
 
As both $t_f - t_2 \gg 1/\Delta E_p$ and $t_1 - t_0  \gg 1/\Delta E_p$, where
$\Delta E_p$ is the energy gap between the nucleon energy $E_p$ and the next
excitation (i.e. the threshold of a nucleon and a pion in the $p$-wave),
the intermediate state contributions from the interpolation fields will be dominated by
the nucleon with the Euclidean propagator $e^{-E_p (t_f -t_0)}$.
From the four-point and two-point functions on the lattice
\begin{eqnarray}  \label{3_pt}
G_{pWp}^{\alpha\beta}& =& \sum_{\vec{x_f}} e^{-i \vec{p}\cdot\vec{x_f}}
 \left\langle\chi_N^{\alpha}(\vec{x_f},t_f) \sum_{\vec{x}} \frac{e^{-i \vec{q}\cdot \vec{x}}}{4\pi} 
 J_{\mu}(\vec{x},t_2)\,J_{\nu}(0,t_1)  \sum_{\vec{x_0}} e^{i \vec{p}\cdot\vec{x_0}}\,\overline{\chi}_N^{\beta}(\vec{x_0}, t_0)\right\rangle , \\
 G_{pp}^{\alpha\beta} &=&  \sum_{\vec{x_f}} e^{-i \vec{p}\cdot\vec{x_f}}\left\langle \chi_N^{\alpha}(\vec{x_f},t_f) \,
 \overline{\chi}_N^{\beta}(\vec{x_0} = 0, t_0)\right\rangle, 
 \end{eqnarray}
we define
\begin{eqnarray}  \label{wmunu_tilde}
\widetilde{W}_{\mu\nu}(\vec{q},\vec{p},\tau) &=&
 \frac{E_p }{m_N} \frac{{\rm Tr} (\Gamma_e G_{pWp})}{{\rm Tr} (\Gamma_e G_{pp})}
 \begin{array}{|l} \\  \\  t _f -t_2 \gg 1/\Delta E_p, \, t_1 - t_0 \gg 1/\Delta E_p \end{array} \nonumber \\
    &=& \frac{E_p }{m_N}\frac{\frac{|Z|^2m_N (E_p + m_N)}{ E_p^2}e^{-E_p(t-t_0)}<N|\sum_{\vec{x}} \frac{e^{-i \vec{q}\cdot \vec{x}}}{4\pi}
  J_{\mu}(\vec{x},t_2) J_{\nu}(0,t_1)|N>}
{\frac{|Z|^2 (E_p +m_N)}{E_p} e^{-E_p(t-t_0)}} \nonumber \\
  &=& <N|\sum_{\vec{x}} \frac{e^{i \vec{q}\cdot \vec{x}}}{4\pi} e^{-i\vec{q}\cdot \vec{x}}
J_{\mu}(\vec{x},\tau) J_{\nu}(0,0)|N>,
\end{eqnarray}
where $\tau = t_2 - t_1$,   $Z$ is the transition matrix element
$\langle 0|\chi_N|N\rangle$, and $\Gamma_e = \frac{1 + \gamma_4}{2}$ is the unpolarized projection to the positive parity nucleon state. 
Inserting intermediate states, 
$\widetilde{W}_{\mu\nu}(\vec{q}^{\,2},\tau)$ becomes
\begin{equation}   \label{wtilde}
\widetilde{W}_{\mu\nu}(\vec{q}^{\,2},\tau)
= \frac{1}{4 \pi}\sum_n \left(\frac{2 m_N}{2 E_n}\right) \delta_{\vec{p}+\vec{q}, \vec{p_n}}\langle N(p)|J_{\mu}(0)|n\rangle
\langle n|J_{\nu}(0) | N(p)\rangle_{spin\,\, ave.} e^{- (E_n - E_p) \tau}.
\end{equation}
Formally, to recover the delta function $\delta(E_n - E_p - \nu)$ in Eq. (\ref{w}) in the continuum formalism, one 
can carry out the inverse Laplace transform with $\tau$ being treated as a dimensionful continuous variable
\begin{equation}  \label{wmunu} 
W_{\mu\nu}(q^2,\nu) = \frac{1}{2m_Ni} \int_{c-i \infty}^{c+i \infty} d\tau\,
e^{\nu\tau} \widetilde{W}_{\mu\nu}(\vec{q}^{\,2}, \tau),
\end{equation} 
with $c > 0$. This is basically doing the anti-Wick rotation back to the 
Minkowski space. 
We will discuss the numerical lattice approach to this conversion from Euclidean space to Minkowski space later. 

 \section{Parton degrees of freedom}  \label{Pdegree}

    In addressing the origin of the Gottfried sum rule violation, it is shown~\cite{Liu:1993cv,Liu:1998um,Liu:1999ak,Liu:2016djw} 
that the contributions to the four-point function of the Euclidean path-integral formulation of 
the hadronic tensor $\widetilde{W}_{\mu\nu}(\vec{q}^{\,2}, \tau)$ in Eq. (\ref{wtilde})  can be classified according 
to different topologies of the quark 
paths between the source and the sink of the proton. Fig. 1(a) and 1(b) represent connected insertions (C.I.) of the
currents.  Here the quark fields from the interpolators $\chi_N$ contract
with that in the currents such that the quark lines flow continuously from $t =0$ to $t =t_f$ and the current insertions are
at $t_1$ and $t_2$. Fig. 1(c), on the other hand, represents a disconnected
insertion (D.I.) where the quark fields from $J_{\mu}$ and $J_{\nu}$
self-contract and, as a consequence, the quark loop is disconnected from the quark paths between
the proton source and sink. Here, ``disconnected'' refers only to the 
quark lines. Of course, quarks propagate in the background gauge fields and all quark paths are ultimately
connected through the gluon field fluctuations.

\begin{figure}[htbp] \label{hadonic_tensor}
\centering
\subfigure[]
{{\includegraphics[width=0.3\hsize]{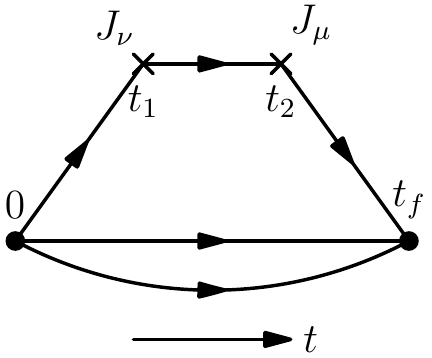}}
  \label{val+CS}}
\subfigure[]
{{\includegraphics[width=0.3\hsize]{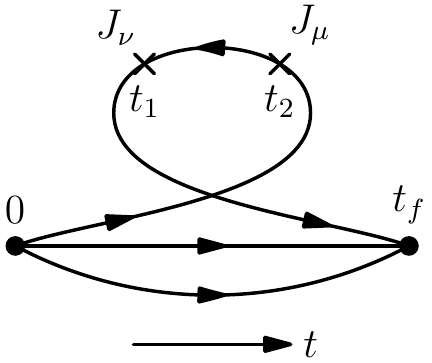}}
  \label{CS}}
\subfigure[]
{{\includegraphics[width=0.3\hsize]{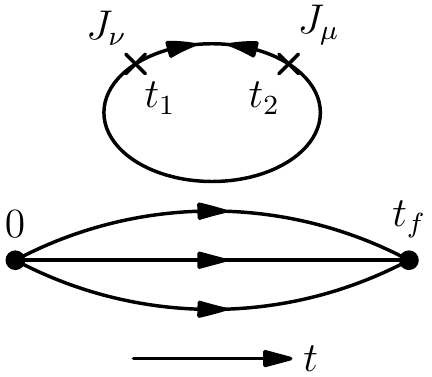}}
 \label{DS}}
\caption{Three gauge invariant and topologically distinct diagrams in the Euclidean-path integral
formulation of the nucleon hadronic tensor. In between the currents
at $t_1$ and $t_2$, the parton degrees of freedom are
  (a) the valence and connected sea (CS) partons $q^{v+cs}$, (b) the CS anti-partons $\bar{q}^{cs}$. Only $u$ and
$d$ are present in (a) and (b) for the nucleon hadronic tensor. (c) the disconnected sea (DS) partons $q^{ds}$ and
anti-partons $\bar{q}^{ds}$ with $q = u, d, s,$ and $c$.}
\end{figure}

We first note that Fig.~\ref{CS}, where the quarks propagate backward in time between $t_1$ and $t_2$
corresponds to the connected sea (CS) anti-partons $\bar{u}^{cs}$ and $\bar{d}^{cs}$, since the quark lines are connected
to the nucleon interpolation fields at $t=0$ and $t =t_f$. This is referred to as `intrinsic bound-valence' 
partons~\cite{Brodsky:1990gn}. By the same token, 
Fig.~\ref{val+CS} gives the valence and CS partons $u^{v + cs}$ and  $d^{v + cs}$. Here the valence 
is defined as 
\begin{equation}  \label{valence}
u^v(d^v)(x) \equiv u^{v + cs}(d^{v + cs})(x) - \bar{u}^{cs}(\bar{d}^{cs})(x),
\end{equation}
with 
\begin{equation}
u^{cs}(x) \equiv \bar{u}^{cs}(x); \hspace{0.5cm} d^{cs}(x) \equiv \bar{d}^{cs}(x).
\end{equation}
On the other hand, Fig.~\ref{DS} gives the disconnected sea (DS) $q^{ds}$ and $\bar{q}^{ds}$ with $\{q = u,d,s,c\}$.  We see that 
while $u$ and $d$ have both CS and DS, strange and charm have only DS. 

     The flavor and valence-sea classification of PDG is summarized in the following Table~\ref{tab:flavor}.

\begin{table}[htbp]  
\caption{Classification of PDF in the nucleon for different flavors. \label{tab:flavor}} 
\vskip\baselineskip
\begin{center}
\begin{tabular}{|c|c|c|c||c|c|c|c|}
\hline
  \multicolumn{4}{|c||}{Valence and Connected Sea} & \multicolumn{4}{|c|}{Disconnected Sea}  \\
\hline
   $u^{v+cs}(x)$ &  $\bar{u}^{cs}(x)$ & $d^{v+cs}(x)$ & $\bar{d}^{cs}(x)$ & $ u^{ds}(x)/ \bar{u}^{ds}(x)$ 
    & $ d^{ds}(x)/ \bar{d}^{ds}(x)$  & $ s(x)/\bar{s}(x)$ & $ c(x)/ \bar{c}(x)$  \\
\hline
 \end{tabular}
\end{center}
\vskip .3cm
\end{table}

It is clear from the path-integral diagrams that there are two sources of the sea partons, one is CS and the other
is DS. In the isospin limit where $\bar{u}^{ds}(x) = \bar{d}^{ds}(x)$, the DS do not contribute to the 
Gottfried sum rule (GSR) violation which reveals that $\int_0^1 dx [\bar{u}(x) - \bar{d}(x)]  < 0$ from DIS 
experiments. The isospin symmetry breaking due to the $u$ and $d$ 
mass difference should be of the order of \mbox{$(m_d - m_u)/m_N$} and cannot explain the large violation of GSR.
Rather, the majority of the violation should come from the CS~\cite{Liu:1993cv}.

\subsection{Small $x$ behavior}

Since the CS parton is in the connected insertion which is flavor non-singlet
like the valence, its small x behavior reflects the leading reggeon exchanges
of $\rho, \omega, a_2...$ and thus should be like
$x^{-1/2}$. On the other hand, the DS is flavor singlet and can have Pomeron exchanges, its small x behavior goes like
$x^{-1}$.  Therefore,  we have 
\begin{eqnarray}
u^{v+cs}(x),\, d^{v+cs}(x), \,\bar{u}_{cs}(x), \,\bar{d}_{cs}(x)_{\,\,\stackrel{\sim}{x \rightarrow 0}} \,\,
x^{-1/2}, \label{x1}\\
 u^{ds}/\bar{u}^{ds}(x), \,  d^{ds}/\bar{d}^{ds}(x),\,  s^{ds}/\bar{s}^{ds}(x)_{\,\,\stackrel{\sim}{x \rightarrow 0}}\,\, x^{-1}.
 \label{x2}
\end{eqnarray}

Since the Gottfried sum rule violation is primarily due to the CS, one expects \\
\mbox{$\bar{u}(x) - \bar{d}(x) = \bar{u}^{cs}(x) - \bar{d}^{cs}(x)$} up to small isospin violation in the DS. Thus, it is not surprising to find that $x (\bar{u}(x) - \bar{d}(x))_{\,\,\stackrel{\longrightarrow}{x \rightarrow 0}} \,\, 0$ in the globally analysis
of PDF~\cite{Lai:2010vv}, the E866 Drell-Yan experiment~\cite{Towell:2001nh}, and the HERMES SIDIS 
experiment~\cite{Ackerstaff:1998sr}.

\subsection{OPE and lattice calculation of moments}  \label{OPE}

\begin{figure}[htbp]
\centering
\subfigure[]
{\includegraphics[width=0.35\hsize]{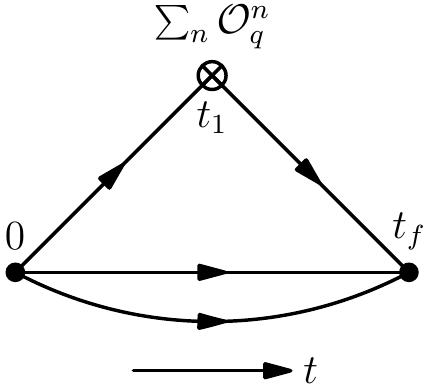}
  \label{CI}}
\hspace*{2cm}
\subfigure[]
{\includegraphics[width=0.35\hsize]{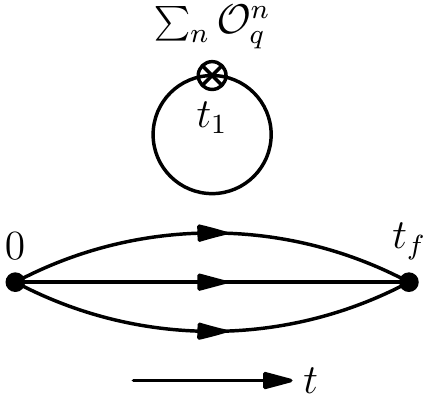}
 \label{DI}}
\caption{The three-point functions after the short-distance expansion of the hadronic tensor from
Fig. 1. (a) The connected insertion (CI) is derived from Fig.~\ref{val+CS} and
Fig.~\ref{CS}. (b) The disconnected insertion (DI) originates from Fig.~\ref{DS}. $\mathcal{O}_q^n$ are
local operators which are the same as derived from OPE.}
\label{fig:DI}
\end{figure}

    Since the fermions are represented by anti-commuting Grassmann numbers, the operator product expansion
(OPE)  entails a short-distance Taylor expansion in the Euclidean path-integral~\cite{Liu:1999ak}. 
Under this short-distance expansion of the hadronic tensor between the current
insertions in the path-integral formalism, Fig.~\ref{val+CS} and Fig.~\ref{CS} become the connected insertions (CI) in
Fig.~\ref{CI} for a series of local operators $\sum_n O_q^n$ in the
three-point functions from which the nucleon matrix elements for the
moments of the CI are obtained. Here the flavor $q = u, d$ are the valence
flavors from the interpolation field. By the same token, the disconnected
four-point functions in Fig.~\ref{DS} become the disconnected insertions (DI)
in Fig.~\ref{DI} for the three-point functions to obtain the DI moments.
Here $q = u, d, s, c$ are the DS flavors in the DI.
The main advantage of the path-integral formalism over the canonical
formalism in Minkowski space is that the parton degrees of freedom are tied to the topology
of the quark skeleton diagrams in Figs.~\ref{val+CS}, ~\ref{CS}, and~\ref{DS}
so that the CS and the DS can be separated.

Lattice QCD can access these three-point functions for the CI and DI which
separately contain the CS and DS and calculations of the moments of the
unpolarized and polarized PDFs for the quarks~\cite{Collins:2017,Bhattacharya:2016zcn,Abdel-Rehim:2015owa,Liu:2017pru} and
glue~\cite{Deka:2013zha,Yang:2016plb,Alexandrou:2016ekb} have been carried out. At the present stage, lattice calculations have reached the physical pion mass point and the systematic errors due to finite volume and finite lattice spacings are beginning to be 
controlled~\cite{Yang:2015uis,Yang:2016plb,Sufian:2016pex}.
However, lattice calculation of the parton moments is limited to a few low moments (about 2 or 3). The higher
moment calculation is impeded by the complication of renormalization and mixing with lower-dimension operators which
leads to power divergences.

\subsection{Separation of CS and DS Partons}   \label{CSDS}

In the global fittings of parton distribution function (PDF), the CS is not separated from the DS and it had
been implicitly assumed that all the anti-partons are from the DS. That's why the GSR violation came as
a surprise. As a result, the fitting has accommodated the $\bar{u}(x) - \bar{d}(x)$ difference from experiment. However,
it is still mostly assumed in the PDF parametrization that the $\bar{u}(x) + \bar{d}(x)$ has the same $x$ dependence as that of 
$s(x) + \bar{s}(x)$. As we discussed above, $\bar{u}(x) + \bar{d}(x) = \bar{u}^{cs}(x) + \bar{d}^{cs}(x) 
+\bar{u}^{ds}(x) + \bar{d}^{ds}(x) $ have both the CS and DS partons and they have different small $x$ behaviors.
This is in contrast to $s(x) + \bar{s}(x)$ where there are only DS partons. An attempt to separate CS from DS anti-partons has
been pursued~\cite{Liu:2012ch}. Combining HERMES data on the strangeness 
parton distribution~\cite{hermes08}, the CT10 global fitting of the $\bar{u}(x) + \bar{d}(x)$ distributions~\cite{Lai:2010vv}, 
and the lattice result of the moment ratio of the strange to $u/d$ in the disconnected insertion, i.e. 
$\langle x\rangle_{s+\bar{s}}/\langle x\rangle_{u+\bar{u}}({\rm DI})$~\cite{doi08}, it is demonstrated ~\cite{Liu:2012ch} 
that the CS and DS partons can be separated and the CS $\bar{u}^{cs}(x)+\bar{d}^{cs}(x)$ distribution of the proton 
is obtained in the region $0.03 < x < 0.4$ at $Q^2 = 2.5 \,\,{\rm GeV}^2$. This assumes that the distribution of 
$\bar{u}^{ds}(x)+\bar{d}^{ds}(x)$ is proportional to that of $s(x) + \bar{s}(x)$, so that the CS partons can be extracted at
$Q^2 = 2.5\,\, {\rm GeV}^2$ through the relation
\begin{equation}  \label{udCS}
\bar{u}^{cs}(x)+\bar{d}^{cs}(x) = \bar{u}(x)+\bar{d}(x) - \frac{1}{R}(s(x) + \bar{s}(x)),
\end{equation}
where $(s(x) + \bar{s}(x))$ is from the HERMES experiment~\cite{hermes08}, $\bar{u}(x)+\bar{d}(x)$ is from
the CT10 gobal fitting of  PDF~\cite{Lai:2010vv}, and $R$ is defined as
  \begin{equation}  \label{ratio}
     R=\frac{\langle x\rangle_{s+\bar{s}}}{\langle x\rangle_{u+\bar{u}}(DI)},
   \end{equation}
and the lattice result $R = 0.857(40)$~\cite{doi08} is used for the extraction. 
\begin{figure}[hbtp]
  \centering
  {{\includegraphics[width=0.45\hsize]{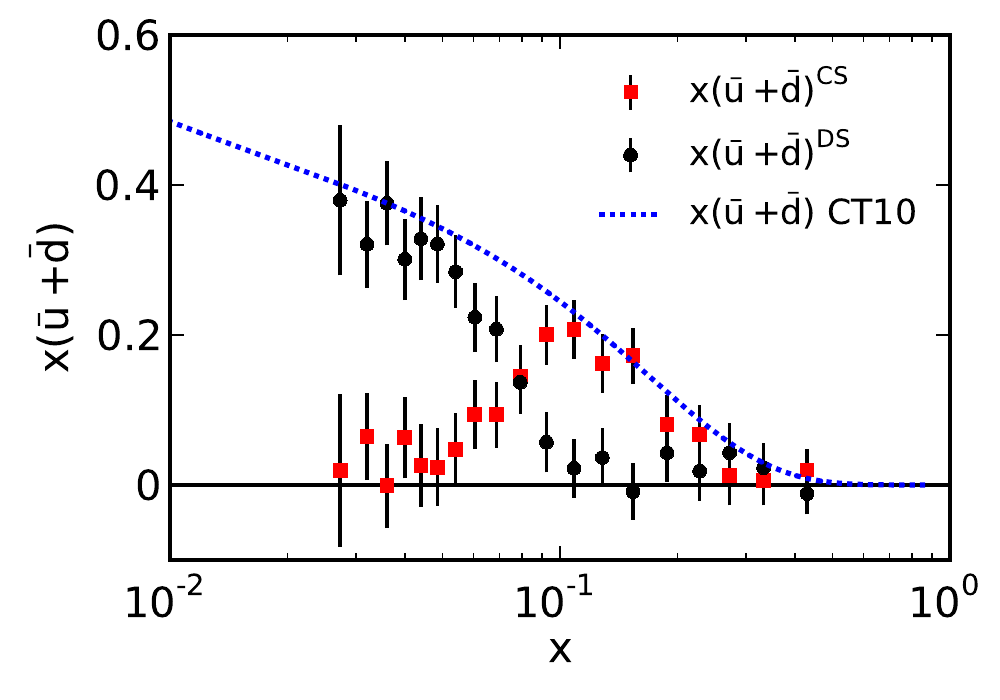}}}
 \hspace{1cm}
  {{\includegraphics[width=0.45\hsize]{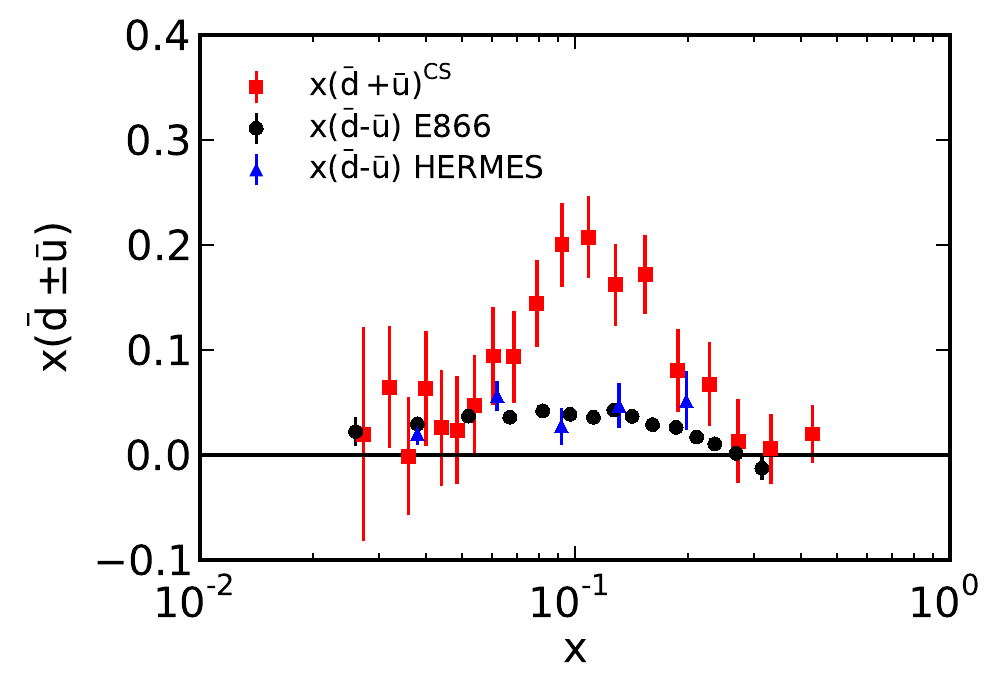}}}
\caption{(Left panel) $x(\bar{u}^{cs}(x) + \bar{d}^{cs}(x))$ obtained from Eq.~(\ref{ratio})  is plotted together with  
$x(\bar{d}(x) + \bar{u}(x))$ from CT10 and $\frac{1}{R}x(s(x)+\bar{s}(x))$ which is taken to be 
$x(u^{ds}(x)+\bar{u}^{ds}(x))$. (Right panel)  $x(\bar{u}^{cs}(x) + \bar{d}^{cs}(x))$ is plotted together with 
$x(\bar{u}^{cs}(x) - \bar{d}^{cs}(x))$ from the E866 and HERMES experiments.}
\label{CSDSu+d}
\end{figure}

The results of $x(\bar{u}(x)+\bar{d}(x) - \frac{1}{R}(s(x) + \bar{s}(x))$, 
$x(\bar{u}^{ds}(x)+\bar{d}^{ds}(x)) = \frac{1}{R}x(s(x) + \bar{s}(x))$
and \\
\mbox{$x(\bar{u}(x)+\bar{d}(x))$} from CT10 at $Q^2 = 2.5\, {\rm GeV^2}$ are plotted in the left panel of 
Fig.~\ref{CSDSu+d}. We see that 
$\bar{u}^{ds}(x)+\bar{d}^{ds}(x)$ is indeed more singular than $\bar{u}^{cs}(x)+\bar{d}^{cs}(x)$  at small $x$ as
expected from Eqs.~(\ref{x1}) and (\ref{x2}). We also plot the extracted $x(\bar{u}^{cs}(x)+\bar{d}^{cs}(x))$ and
$x(\bar{u}^{cs}(x)-\bar{d}^{cs}(x))$ from E866 Drell-Yan experiment~\cite{Towell:2001nh} and HERMES SIDIS 
experiment~\cite{Ackerstaff:1998sr}. We see that they are in the same $x$-range and peak around $x = 0.1$.
It should be pointed out that the CS partons from Eq.~(\ref{udCS}) was based on the HERMES data in 2008~\cite{hermes08}. 
These results will be updated with the 2014 HERMES data~\cite{Airapetian:2013zaw} and the lattice result of $R$
in Eq.~(\ref{ratio}) at the physical pion point and with the associated systematic errors on infinite volume and continuum limits
taken into account~\cite{Sun:2015pea,Sun17}. Since the new HERMES data on 
\mbox{$x(\bar{s}(x)+\bar{s}(x))$}~\cite{Airapetian:2013zaw} are generally smaller than those of the 2008 data~\cite{hermes08} in the
range of $0.03 < x < 0.4$ and if the new lattice value of $R$ is not too far from the one~\cite{doi08} used to extract the CS partons shown in Fig.~\ref{CSDSu+d}, the to-be-updated CS partons are expected to be more prominent in this range of $x$. The results of the CS partons will change somewhat, but their qualitative features are expected to remain.

\subsection{Lattice calculation of PDF}  \label{lattice-PDF}

The extraction of  $\bar{u}^{cs}(x)+\bar{d}^{cs}(x)$ in Eq.~(\ref{udCS}) is based on the assumption
that the distribution of $s(x) + \bar{s}(x)$ is proportional to that of $u^{ds}(x) + \bar{u}^{ds}(x)$ or
$d^{ds}(x) + \bar{d}^{ds}(x)$ so that their ratio can be obtained via the ratio $R$ in Eq.~(\ref{ratio}).
It would be better to calculate  $\widetilde{W}_{\mu\nu}$ represented in Figs.~\ref{val+CS}, \ref{CS} and
\ref{DS} directly on the lattice. However, there is a numerical complication in that an inverse Laplace transform is
involved in converting $\widetilde{W}_{\mu\nu}$ to $W_{\mu\nu}$ in Minkowski space as in Eq.~(\ref{wmunu}).
An improved Maximum Entropy Method (MEM)~\cite{Burnier:2013nla} which can lead to more stable fit is proposed to solve this 
inverse problem~\cite{Liu:2016djw}. Recently, there is another approach to calculating PDF on the lattice via the
quasi-PDF~\cite{Ji:2013dva,Lin:2014zya,Alexandrou:2015rja,Chen:2016utp} in the large momentum frame.
Both approaches are at their infancy and still face many numerical challenges. They are not as mature as the
lattice calculation of moments and matrix elements which are at the stage of finalizing the calculations with all the
systematic errors under consideration.

\section{NNLO evolution equations for the valence, CS, and DS}  \label{CSDSEE}

   We see from Sec.~\ref{OPE}, that, under the short-distance expansion of the hadronic tensor,  
the connected four-point functions in Fig.~\ref{val+CS} and Fig.~\ref{CS} become the connected insertions (CI) in
Fig.~\ref{CI} for a series of local operators $\sum_n O_q^n$ in the
three-point functions from which the nucleon matrix elements for the
moments of the CI are obtained. Here the flavor $q = u, d$ are the valence
flavors from the interpolation field. By the same token, the disconnected
four-point functions in Fig.~\ref{DS} become the disconnected insertions (DI)
in Fig.~\ref{DI} for the three-point functions to obtain the DI moments. It is clear from the operator analysis
of operator scaling and mixing, only the DI can mix with the glue operator. Since the quark lines in the CI
are connected between the current operators and the interpolation fields of the nucleon source and sink , it does not
have the annihilation channel to mix with glue operators. 
As a consequence, one deduces that the CS evolves the same way as the valence, i.e. they can evolve into
valence, CS, and DS. But the DS cannot evolve into valence and CS. On the other hand, the gluons can split into DS, but 
not into the valence and CS, since their operators do not mix. 


    With the above operator analysis, it is straightforward to write down the extended NNLO DGLAP evolution equations which
 accommodates the separately evolved CS and DS.

\begin{eqnarray}  
\frac{dq_i^{v+cs}}{dt} \!\! &=& \!\! P_{ii}^{c} \otimes q_i^{v+cs} + P_{i\bar{i}}^{c} \otimes \bar{q}_i^{cs};  \label{v+cs} \\
\frac{d\bar{q}_i^{cs}}{dt}  \!\! &=& \!\!  P_{\bar{i}\bar{i}}^{c} \otimes \bar{q}_i^{cs} + P_{\bar{i}i}^{c} \otimes q_i^{v+cs}; \label{csbar}  \\
\frac{dq_i^{ds}}{dt}  \!\! &=& \!\!  \sum_k P_{ik}^{cd} \otimes q_k^{ds} + \sum_k P_{i\bar{k}}^{cd} \otimes \bar{q}_k^{ds} 
                   + \sum_k  P_{ik}^{d} \otimes q_k^{v+cs}
                   +  \sum_k  P_{i\bar{k}}^{d} \otimes \bar{q}_k^{cs} + P_{ig} \otimes g;   \label{ds}\\
\frac{d\bar{q}_i^{ds}}{dt}  \!\! &=& \!\!  \sum_k P_{\bar{i}\bar{k}}^{cd} \otimes \bar{q}_k^{ds} 
                         + \sum_k P_{\bar{i}k}^{cd} \otimes q_k^{ds} + \sum_k  P_{\bar{i}k}^{d} \otimes q_k^{v+cs}
                   +  \sum_k  P_{\bar{i}\bar{k}}^{d} \otimes \bar{q}_k^{cs} + P_{ig} \otimes g;   \label{dsbar}   \\
\frac{dg}{dt}  \!\! &=& \!\! \sum_k (P_{gk} \otimes (q_k^{v+cs}+ q_k^{ds}) + P_{g\bar{k}}\otimes (\bar{q}_k^{cs}+\bar{q}_k^{ds}))   + P_{gg} \otimes g   \label{new-g},
\end{eqnarray}
where $P_{ii}^{c} = P_{\bar{i}\bar{i}}^{c},  P_{i\bar{i}}^{c} =  P_{\bar{i}i}^{c}$ and they involve only 
connected diagrams where the quark line is connected between the initial quark and the pinched current point 
(e.g. Fig. 2a and the left most one in 2b in Ref.~\cite{vog96}). $P^{d}$, on the other hand,  involves 
only the quark-line disconnected diagrams where the pinched point is on the quarks/antiquarks in the loop 
(e.g. two right diagrams in Fig. 2b in Ref.~\cite{vog96}). Note that in NNLO, there is three-gluon exchange between
the quark loop with current insertions and the quark line from the nucleon (both valence and DS) as illustrated in
Fig. 1(b) in Ref.~\cite{lv91}. This implies $P_{ik}^{d} = P_{\bar{i}\bar{k}}^{d} \neq P_{i\bar{k}}^{d} =  P_{\bar{i}k}^{d}$. 
Thus in NNLO, the evolution itself can induce $q_i^{ds} \neq \bar{q}_i^{ds}$ by the valence and the DS. 
$P_{ik}^{cd} = P_{\bar{i}\bar{k}}^{cd}$ and $P_{\bar{i}k}^{cd} = P_{i\bar{k}}^{cd}$ involve evolutions 
from DS to DS and they have both connected and disconnected diagrams, i.e.
\begin{equation}  \label{Pcd}
P_{ik}^{cd} = P_{ii}^{c}\delta_{ik} + P_{ik}^{d}.
\end{equation}

     We shall compare these equations to Eqs.~(\ref{sigma_v},\ref{q+},\ref{q-},\ref{sigma_sum},\ref{glue}). We first note
that the quantities defined in Eqs.~(\ref{def_sigma}) and (\ref{def_qpm}) have the following comnponents
   
\begin{eqnarray}
q_i^- &\equiv & q_i - \bar{q}_i = q_i^{v+cs} - \bar{q}_i^{cs} + q_i^{ds}(x) - \bar{q}_i^{ds}(x);  \label{def_q-}\\
\Sigma &\equiv & \sum_{i} (q_i + \bar{q}_i) = \sum_{i=u,d} ( q_i^{v+cs} + \bar{q}_i^{cs}) + \sum_{i=u,d,s} (q_i^{ds} + \bar{q}_i^{ds}); \\
q_i^+ &\equiv & q_i + \bar{q}_i - \frac{1}{N_f} \Sigma = \left\{ \begin{array}{ll}
                          q_i^{v+cs} + \bar{q}_i^{cs} + q_i^{ds} + \bar{q}_i^{ds} - \frac{1}{N_f}\Sigma & \mbox{i = u,d}; \\
                          s + \bar{s}  - \frac{1}{N_f}\Sigma & \mbox{i = s}
                          \end{array}  \right.
 \end{eqnarray}

Taking the combination Eq.~(\ref{v+cs}) - Eq.~(\ref{csbar}) + Eq.~(\ref{ds}) - Eq.~(\ref{dsbar}), we have
\begin{equation}   \label{q-_new}
\frac{dq_i^-}{dt}=  P_{qq}^- \otimes q_i^- + P_{ds}^- \otimes \sum_k q_k^-, 
\end{equation}
with
\begin{equation}
P_{qq}^- = P_{qq}^{c} - P_{q\bar{q}}^{c} \equiv P_{qq}^{v} - P_{q\bar{q}}^{v}; \,\,\,  
P_{ds}^- = P_{qq}^d - P_{q\bar{q}}^d \equiv P_{qq}^s - P_{q\bar{q}}^s.
\end{equation}
This is just Eq.~(\ref{q-}) with the inhomogeneous term being the sum of $q_k^-$.
The first term in Eq.~(\ref{q-_new}) is from the difference of Eqs.~(\ref{v+cs}) and (\ref{csbar})
and the flavor-diagonal parts ($\delta_{ik}$) of the first two terms in Eqs.~(\ref{ds}) and (\ref{dsbar}); 
while the second term is from the rest of Eqs.~(\ref{ds}) and (\ref{dsbar}). Thus, we now understand
that Eq.~(\ref{q-}) is the sum of the evolution of $q_i^{v+cs} - \bar{q}_i^{cs}$ and 
$q_i^{ds} - \bar{q}_i^{ds}$. Note the inhomogeneous term only enters in NNLO where 
$P_{ik}^{d} = P_{\bar{i}\bar{k}}^{d} \neq P_{i\bar{k}}^{d} =  P_{\bar{i}k}^{d}$.
It is clear now that $q^-(x)$ is not the valence, as discussed in Sec.~\ref{EE}, it includes $ q_i^{ds}(x) - \bar{q}_i^{ds}(x)$.
in Eq.~(\ref{def_q-}). The proper definition of the valence is Eq.~(\ref{valence}). Eq.~(\ref{sigma_v}) is simply the sum of  Eq.~(\ref{q-_new}) over flavor. 

Utilizing Eq.~(\ref{Pcd}), the equation for $\Sigma$ is
\begin{eqnarray}  
\frac{d\Sigma}{dt} 
         &=& P_{qq}^+ \otimes ((\sum_{i =1}^2 q_i^{v+cs} + \bar{q}_i^{cs}) + \sum_k (q_k^{ds} + \bar{q}_k^{ds})) \nonumber \\
         &+& \sum_{i,k} (P_{ik}^{d} + P_{\bar{i}k}^{d}) \otimes (q_k^{v+cs} + \bar{q}_k^{cs} + q_k^{ds} + \bar{q}_k^{ds})
       + 2 \sum_i P_{ig} \otimes g.
\end{eqnarray}
  This can be written in terms of $\Sigma$
\begin{equation} \label{sum_new}
\frac{d\Sigma}{dt}=  P_{\Sigma\Sigma} \otimes \Sigma + P_{\Sigma g} \otimes g, 
\end{equation}
with 
\begin{eqnarray}  \label{Psum_new}
P_{\Sigma\Sigma} &=& P_{qq}^{c} + P_{q\bar{q}}^{c} + N_f (P_{qq}^{d} + P_{\bar{q}q}^{d}); \\
P_{\Sigma g} &=& 2 N_f P_{qg}.
\end{eqnarray}
Given $P_{qq}^c \equiv P_{qq}^v$ and $P_{qq}^d \equiv P_{qq}^s$, Eq.~(\ref{sum_new}) is just Eq.~(\ref{sigma_sum}). 

Similarly, one can show that the equation for $q^+$ has the following form for $i = u,d$ and $s$
\begin{equation}   \label{q+_new}
\frac{dq_i^+}{dt} =  P_{qq}^+ \otimes q_i^+,
\end{equation}
which is the same as in Eq.~(\ref{q+}) where 
$P_{qq}^+ = P_{qq}^{c} + P_{q\bar{q}}^{c} \equiv P_{qq}^{v} + P_{q\bar{q}}^{v}$.
Finally, Eq.~(\ref{new-g}) is just Eq.~(\ref{glue}) with $P_{g\Sigma} = P_{gq} = P_{g\bar{q}}$.

\subsection{Comments}     

     Now that the extended evolution equations are derived, several comments are in order.

\begin{itemize}

\item 
    Due to the linear nature of the DGLAP equations,  the 9 equations in 
Eqs.~(\ref{sigma_v},\ref{q+},\ref{q-},\ref{sigma_sum},\ref{glue}) can be obtained from the linear combinations of the extended
11 evolutions equations in Eqs.~(\ref{v+cs}, \ref{csbar}, \ref{ds}, \ref{dsbar}, \ref{new-g}). The two extra equations is to
accommodate the CS partons for the the $u$ and $d$ flavors.  These extended equations are ready to accommodate  the most general case with $s \neq \bar{s}, u^{ds} \neq \bar{u}^{ds}, d^{ds} \neq \bar{d}^{ds}$ in addition to flavor dependent DS.

\item
    The valence is defined as $q_i^v \equiv q_i^{v+cs} - \bar{q}_i^{cs}$ which is not the same as $q_i^-$ unless
$q_i^{ds} = \bar{q}_i^{ds}$. This alleviates the potential confusion that strange partons are part of the valence when 
$s(x) \neq \bar{s}(x)$.

\item If one does not distinguish CS from DS, the usual DGLAP equations in Eqs.~(\ref{sigma_v},\ref{q+},\ref{q-},\ref{sigma_sum},\ref{glue}) are adequate. Why does one need to extend them to have separate CS and DS? One of the major reasons is to be
able to compare with lattice calculation and, in some cases, they can be used to help constrain the global PDF analysis. 
As we explained in Sec.~\ref{lattice-PDF}, the lattice calculations of nucleon 
matrix elements are mature with all the systematic errors taken into account. They are ready to produce results which can confront
experiments. However, the lattice calculation are organized in terms of CI in Fig.~\ref{CI} which  are the moments for the valence and
CS partons and DI in Fig.~\ref{DI} which are for the DS partons. On the other hand, the current global fittings of PDF do have
the valence separated, but the CS and DS are lumped together as the total sea. Consequently, no direct comparison
can be made between the lattice moments and those of PDF, except for a few quantities such as $\langle x \rangle_{u-d}$
and $\langle x\rangle_s$.

\item The need to separate CS from DS is particularly acute in the polarized PDF where much interest is focused
on the quark and glue spins, and their orbital angular momenta. To address the `proton spin crisis' where the quark spin is found to
contribute only $\sim 30\%$ of the proton spin, the lattice calculation can be carried out for the flavor-singlet axial-vector current matrix elements in the CI and DI. Lattice calculations~\cite{Dong:1995rx,Fukugita:1994fh,Gusken:1999as,QCDSF:2011aa,Babich:2010at,Engelhardt:2012gd,Abdel-Rehim:2013wlz,Chambers:2015bka} have shown that the the matrix element from the DI of the flavor-singlet axial-vector current is negative. This reduces that from the CI to make the total quark spin smaller than expected from the valence contribution. Further examination of the negative DI contribution can be understood in terms of the cancellation between the pseudoscalar density term and the anomaly term through the anomalous Ward identity~\cite{Gong:2015iir}. One would like to compare these findings to experiments. But this is not attainable unless and until the global fitting manages to separate the CS from the DS in polarized DIS and Drell-Yan processes. 
 
\item An example is given to separate CS from DS in Sec.~\ref{CSDS} which utilizes the combined global PDF, experimental
data and lattice calculation to do the job. This is done for one $Q^2$. Only through the fully separated CS and DS degrees of 
freedom in the extended evolutions can the CS and DS remain separated at different $Q^2$. This aspect is essential for the
global analysis of PDF with fully separated CS and DS as a function of both $x$ and $Q^2$.


 %
 


\end{itemize}

\section{Summary}  \label{sum}

     The roles of the connected-sea (CS) and  disconnected-sea (DS) partons, as revealed in the path-integral formulation of the hadronic tensor in the Euclidean space, are clarified in terms of the Gottfried sum rule violation, their small $x$ behaviors, the moments of PDF, and evolution. An example is given to show how the CS can be separated from DS by combining the CT10 PDF, HERMES SIDIS data on the strange parton distribution and the lattice calculation of the ratio of the second moment of the strange vs the $u/d$ in the DI.     
    
     From the short-distance expansion which is equivalent to OPE in Minkowski space, it is shown that the valence and CS partons
 merge in the moments of the connected insertion (CI), while the DS goes into the moments of the disconnected insertion (DI). 
 Since only the DI mixes with the glue operators, it implies that the CS evolves the same way as the valence. 
 The extended DGLAP equations are thus derived which entails separate equations for the CS and DS. Upon linear combinations, it
 is shown that they reproduce the conventional DGLAP equations where the CS and DS are not separated.
     
     Special emphasis is placed on the need to have separately evolved CS and DS so that comparison with lattice calculations of
unpolarized and polarized moments of PDF can be made. Only with the extended DGLAP equations will the CS and DS 
remain separated at different $Q^2$ to facilitate global fitting of PDF with separated CS and DS partons.

\section{Acknowledgment}
 This work is partially support by the U.S. DOE grant DE-SC0013065.
The author is indebted to Pavel Nadolsky for his notes and T.J. Hou, J.C. Peng, W.K. Tung and C.-P. Yuan for many 
insightful discussions.


\end{document}